# Graphene Electric Double-Layer Transistors for Enhanced-Sensitivity Label-Free Detection of Human Serum Albumin

Arslan Liaquat[1], Ghassem Baridi[2], Federico Rapuzzi[1], Daniele Goldoni[2], Vito Clericò[3], El Hadj Abidi[3], Yahya Moubarak Meziani[3], Mario Amado[3], Enrique Diez[3], Naveen Kumar[1],Camilia Coletti[4], Beatrice Cipriani[5], Hender Lopez[5], Giorgia Brancolini[6], Leonardo Martini[1], Luigi Rovati[2], Francesco Rossella[1*]

[1] Dipartimento di Scienze Fisiche, Informatiche e Matematiche, University of Modena and Reggio Emilia, Via Campi 213/A, 41125 Modena, Italy

[2] Department of Engineering Enzo Ferrari, University of Modena and Reggio Emilia, via Vignolese 905, 41125 Modena, Italy

[3]USAL–Nanolab, Departamento de Física Fundamental, University of Salamanca, Salamanca 37008, Spain

[4] Center for Nanotechnology Innovation @NEST, Istituto Italiano di Tecnologia, Piazza San Silvestro 12, I-56127 Pisa, Italy

[5] School of Physics, Clinical and Optometric Sciences, Technological University

Dublin, Grangegorman, Ireland

[6] Institute Nanoscience, CNR, S3, Via G. Campi 213/A, Modena

Corresponding author: Arslan Liaquat, arslan.liaquat@unimore.it;
Francesco Rossella, francesco.rossella@unimore.it

## Abstract

Accurate and sensitive detection of human serum albumin (HSA) is essential for the early diagnosis and monitoring of renal and hepatic disorders. Here, we present a graphene-based electrolyte-gated field-effect transistor (EGFET) for label-free, real-time quantification of HSA under non-Faradaic operation. The device exploits the high interfacial capacitance of the electric double layer (EDL) to transduce electrostatic perturbations induced by albumin adsorption into measurable conductance modulation. Negatively charged HSA molecules induce systematic modulation of the graphene channel, producing a concentration-dependent displacement of the Dirac voltage consistent with p-type doping.

To establish a molecular-level interpretation of the sensing response, Brownian Dynamics simulations show that HSA adsorbs onto graphene through multiple adsorption orientations associated with heterogeneous interfacial charge distributions and variable dipole alignments relative to the surface. Adsorption is energetically stabilized primarily by van der Waals interactions.

Systematic analysis of transfer characteristics across concentrations ranging from 0.01 to 30mg.mL$^{-1}$ reveals a clear correlation between surface charge density and carrier transport modulation within the electric double layer. The optimized device exhibits a limit of detection of 0.0087 mg mL$^{-1}$ and a linear dynamic range extending to 10 mg mL$^{-1}$. The response remains non-Faradaic under sub-volt operation with reversible and reproducible behavior. The use of an inverse-mobility analytical metric highlights the role of disorder-enhanced carrier scattering in signal amplification, enabling sensitive electrostatic detection while preserving reversible device operation. These results establish liquid-gated graphene EGFETs as a promising platform for quantitative protein sensing and provide insight into disorder-mediated transport mechanisms in graphene bioelectronic devices.

## 1. Introduction

Human serum albumin (HSA) is the predominant plasma protein, accounting for nearly 60 % of total serum protein in healthy adults and performing essential roles in oncotic-pressure regulation, pH buffering, and transport of numerous ligands[1-3]. Even modest deviations from its physiological range signal pathophysiological stress: elevated urinary albumin (micro-albuminuria) marks early glomerular injury and renal dysfunction, while hypoalbuminemia accompanies hepatic failure, systemic inflammation, and malnutrition [4-6]. Reliable, quantitative determination of HSA is therefore vital for early diagnosis and disease monitoring.

Conventional analytical methods—including dye-binding assays, enzyme-linked immunosorbent assays (ELISA), and high-performance liquid chromatography (HPLC) remain clinical standards because of their specificity and reproducibility[7-9]. Yet these approaches require complex sample preparation, costly instrumentation, and trained personnel, and they generally reach limits of detection (LOD) in the microgram-per-milliliter range[7] insufficient for early-stage diagnostics. The drive to detect albumin at nanogram levels has stimulated the development of more sensitive, portable detection platforms.

Nanomaterial-based optical probes have significantly improved protein detection sensitivity by combining molecular recognition with photonic signal amplification. Metallic and semiconductor nanostructures enable highly sensitive colorimetric and fluorescent assays, achieving nanogram-level detection in some cases [10-14]. However, optical techniques often require fluorescent labels, excitation optics, and complex instrumentation, and may suffer from photobleaching, matrix interference, and limited real-time capability[15]. These limitations have motivated the development of alternative sensing strategies based on direct electrical transduction.

The emergence of two-dimensional (2D) nano-materials has opened a complementary route based on direct electrical transduction. Graphene, with atomic thickness, high carrier mobility, and chemical robustness, can register minute electrostatic perturbations caused by molecular adsorption[16]. When operated under electrolyte gating, a nanometer-scale electric double layer (EDL) forms at the graphene–electrolyte interface, generating gate capacitances several orders of magnitude higher than conventional oxide dielectrics [17-19]. Such strong capacitive coupling enables efficient carrier modulation at sub-volt bias, amplifying the influence of individual adsorbates and allowing label-free detection of biomolecules at ultralow concentrations. Within the 2D-materials family, graphene remains the prototypical platform for solution-gated transistors [20].

Despite these advantages, converting graphene's intrinsic sensitivity into quantitatively interpretable biosensing remains challenging. Many GFETs exhibit drift, hysteresis, and ionic-screening effects in physiological electrolytes, obscuring the relation between adsorption dynamics and electrical response [21-22]. Adsorbed water and oxygen further cause charge-noise fluctuations and time-dependent doping, producing baseline instability [23-24]. Advances in electrolyte composition and cation pre-doping have reduced drift and improved stability [25].

Refinements in device architecture, electrolyte formulation, and surface functionalization have further improved analytical fidelity [27-29]. For example, Ohno et al. and Hess et al. elucidated how surface-potential fluctuations and ionic screening govern sensitivity in electrolyte environments,[27-28] while Kwon et al. demonstrated adsorption-induced modulation of screening in polymer-functionalized graphene transistors[29].Recently, Kang et al. reported a real-time, label-free graphene EGFET achieving sub-milligram-per-milliliter sensitivity and high reproducibility.[8]

These advances collectively mark the evolution of graphene FET biosensors from qualitative demonstrations to quantitative analytical tools.

While most reported graphene biosensors primarily interpret sensing response through Dirac voltage shifts, the role of adsorption-induced disorder and mobility modulation remains comparatively underexplored. In this work, we address this limitation by introducing an inverse-mobility analytical framework that captures adsorption-induced carrier scattering in graphene, enabling a disorder-mediated sensing mechanism that complements conventional Dirac-point-based detection. To meet these analytical and clinical challenges, we present a graphene-based Electrolyte-Gated Field-Effect Transistor (EGFET) for label-free, real-time, highly sensitive detection of human serum albumin. Leveraging graphene's high carrier mobility and the strong capacitive coupling of the EDL, the device translates nanoscale electrostatic perturbations from albumin adsorption into measurable conductance variations. Systematic analysis of Dirac-point modulation, carrier mobility, and disorder-related transport broadening contact-resistance evolution reveals a non-Faradaic, capacitive electrostatic mechanism underpinning detection. To analyze charge-transport behavior quantitatively, this work employs the Fitting Transfer Model (FTM) introduced by Kim et al., which separates contact and channel resistances and extracts carrier mobility and residual charge density from transfer curves [26]. The sensor achieves a limit of detection 0.0087 mg $mL^{-1}$ (ultra trace range), comparable to advanced optical nanoprobes yet while maintaining reproducible and stable operation. Beyond early-stage albumin diagnostics, this study establishes a mechanistic framework linking interfacial electrostatics to charge transport in graphene, paving the way for next-generation, low-power, multiplexable point-of-care biosensing technologies.

Beyond early-stage diagnostics, quantitative monitoring of trace albumin concentrations is also important in dialysis applications. In clinical practice, precise quantification of trace albumin in dialysate has crucial diagnostic value. Albumin loss during hemodialysis depends on membrane type and treatment modality: conventional low-flux membranes restrict protein passage, whereas high-flux (HF) and high-cutoff (HCO) membranes permit measurable leakage into the spent dialysate [30-31]. Although typically small, cumulative albumin depletion can contribute to hypoalbuminemia and impaired nutritional status[32]. Optimization of membrane selectivity thus requires real-time assessment of albumin transport across the dialysis barrier. Medium- and high-cutoff membranes improve clearance of middle-molecular-weight uremic toxins while maintaining albumin loss below ~3 g per session [33]. Nevertheless, trace albumin concentrations are consistently detected in spent dialysate commonly within 0.02–1 mg $mL^{-1}$, depending on membrane porosity, operating pressure, and treatment duration [34]. Monitoring such ultralow levels is essential both for evaluating membrane performance and for assessing protein-energy balance in long-term dialysis patients.

## 2. Experimental Section

### 2.1 Materials

All reagents were of analytical grade and used as received without further purification. Human Serum Albumin (HSA) (200 $mg\cdot mL^{-1}$; Grifols, Spain) was used as the analyte. Ultrapure water (resistivity ≥ 18.2 MΩ·cm; Sigma-Aldrich) was employed for all dilutions and rinsing. Acetone (≥99.5%, Merck) and isopropanol (IPA) (≥99.8%, Merck) were used for cleaning the graphene chips, and high-purity nitrogen gas (99.999%) was used for drying and removing residual moisture.

Monolayer graphene field-effect transistor (GFET) chips were purchased from Graphenea S.A. (San Sebastián, Spain). Each device consisted of a single-layer graphene channel on a $Si/SiO_2$ substrate (90 nm oxide) with gold source and drain electrodes (50 nm). Devices were used without surface functionalization to preserve the intrinsic graphene-electrolyte interface in their pristine form to preserve the intrinsic graphene–electrolyte interface and ensure that sensing response arose from electrostatic interactions rather than chemical modifications.

Electrical transport measurements were conducted using a Keithley 2612 SourceMeter controlled through Ni-VISA software. Calibrated Gilson Pipetman micropipettes (P200, P1000) with low-protein-binding sterile tips were used for all pipetting operations. Albumin dilutions were stored in labeled 1.5 mL low-binding Eppendorf microtubes (Eppendorf AG, Germany). All measurements were conducted in a clean, low-humidity environment (22–25 °C). Data was processed using OriginPro 2025 (OriginLab).

### 2.2 Preparation of Human Serum Albumin (HSA) Dilution Series

A 200 mg·mL$^{-1}$ HSA stock solution was equilibrated to room temperature and gently inverted several times to ensure complete homogenization. Serial dilutions were prepared in ultrapure water (18.2 MΩ·cm) using a Gilson Pipetman P1000 to obtain final HSA concentrations of 0.01, 0.05, 0.1, 0.5, 1, 5, 10 and 30 mg·mL$^{-1}$. Each dilution was prepared in a pre-labelled low-binding microtube, vortexed gently for 5 s, and briefly centrifuged (1,000 rpm, 10 s) to eliminate air bubbles. The solutions were inspected visually for transparency and absence of aggregates, and representative samples were checked for pH stability (6.8–7.4). Dilutions were stored at 4 °C and equilibrated to room temperature before each measurement.

Note that the commercial 20% HSA stock formulation contains physiological saline. Therefore, serial dilution in ultrapure water results in progressively lower ionic strength across the concentration series, which can influence electrostatic screening length. This variation is addressed in the parameter extraction model through the use of an effective gate capacitance parameter.

### 2.3 Rigid Body Docking and Dipole Orientation

To relate protein adsorption to the device's electrostatic response, rigid docking configurations from Brownian Dynamics (BD) simulations were analyzed to determine the orientation of the Human Serum Albumin (HSA) molecular dipole relative to a graphene surface. The 3D structure of HSA (585 amino acids) was obtained from the AlphaFold Protein Structure Database (ID: AF-F6KPG5-F1-v4) [47]. Protonation states at pH 7.4 were assigned using PDB2PQR and PROPKA[48,49], yielding a net negative charge of -13, and atomic charges were defined using the OPLS force field[50]. The graphene surface was modeled as a 145 × 145 Å bilayer of neutral Lennard-Jones carbon atoms.

Protein–surface docking was performed using SDA 7[51], treating both protein and the graphene surface as rigid bodies. A total of 5000 BD trajectories were generated, with the protein undergoing stochastic translation and rotation along the z-axis while the graphene remained fixed, according to a previously established protocol[52]. Simulations used an implicit solvent force field[53], with electrostatic and desolvation parameters as previously described[54]. Each run started with a 60 Å separation between centers of mass of the protein and the graphene surface. Diffusion coefficients (0.00611 Å²/ps translational and $3.13 \times 10^{-5}$ rad²/ps rotational) were obtained from HydroPro[55].

From all trajectories, 2000 docked configurations were retained and clustered using SDA's single-linkage method in two stages based on the Cα RMSD (3 Å cutoff).

### 2.4 Graphene EGFET Electrical Characterization and Parameter Extraction

Prior to each measurement, GFET devices were annealed on a hot plate at 110 °C for 10 min to remove residual adsorbates and moisture and then cleaned sequentially in thermalized acetone and isopropanol for 3 min each and dried with nitrogen.

A 5 µL droplet of the prepared HSA solution was carefully drop-cast onto the graphene channel, ensuring complete coverage between the source and drain electrodes while avoiding the metallic pads. The droplet was left undisturbed for 5 minutes to allow albumin adsorption and electric double layer (EDL) stabilization at the graphene-electrolyte interface.

Two electrical measurement modes were employed: first is the Output characteristics (IV curves), where the drain–source current Ids was recorded as a function of drain–source voltage $V_{ds}$ sweep, at source-gate voltage $V_{tg}$ = 0 V to assess channel conductance and linearity; second recorded curve is Transfer Curve measured at $V_{ds}$ = 200 mV while sweeping the top-gate voltage $V_{tg}$ from 0 to 2 V to capture ambipolar behavior and locate the Dirac point position. After each run, devices were rinsed thoroughly with Acetone and isopropanol and dried with nitrogen to remove unbound proteins. Each HSA concentration was tested in triplicate to evaluate reproducibility and minimize statistical error.

The carrier transport in the graphene channel was analyzed using a field-effect transport model commonly applied to graphene transistors. In this framework, the channel conductivity is expressed as:

$$\sigma = e\mu\sqrt{{n_o}^2 + n^2}$$

Where $e$ is the elementary charge, $\mu$ is the carrier mobility, $n$ is the gate-induced carrier density, and $n_o$ is the residual carrier density at the charge-neutrality point, accounting for disorder-induced charge puddles and local electrostatic inhomogeneity. The gate-induced carrier density was determined from the applied gate voltage through the effective gate capacitance Ceff.

$$n = \frac{C_{eff}(V_g - V_{Dirac})}{e}$$

In the present analysis, the mobility was extracted from the hole branch of the transfer characteristics and is therefore denoted as μh. Although the transport model assumes a single carrier mobility, the parameter is obtained from the hole side of the transfer curve to avoid asymmetry effects near the electron branch.

Electrical transport data were analyzed as a function of HSA concentration to quantify the sensing response and extract key physical parameters. The primary analytical metric used in this work is the inverse mobility signal ($S_{\mu}$), defined as: $S\mu = \frac{1}{\mu_h} \times 10^4$

Where $\mu_h$ is the hole mobility extracted from transfer characteristics. To extract transport parameters, including carrier mobility (μ), contact resistance ($R_c$), and residual carrier density ($n_0$), transfer curves were fitted using a nonlinear resistance model adapted for liquid-gated graphene transistors. In this framework, the total measured device resistance is expressed as the sum of contact and channel contributions:

$$R = Rc + \frac{L}{W} . \frac{1}{\mu \, C_{eff}\sqrt{no^2 + (Vg - Vdirac)^2}}$$

Where L and W are the channel length and width, respectively; 'Vg' is the gate voltage; and 'VDirac' is the charge neutrality point. Additional details of the fitting procedure and parameter extraction are provided in the Supplementary Information (Section S3).The effective gate capacitance 'Ceff' was treated as a fixed fitting parameter and set to 1.65 μF cm$^{-2}$, selected within the expected range for low-ionic-strength electrolyte-gated graphene devices (~1–2 μF cm$^{-2}$) to ensure stable and physically consistent fitting across the full concentration series. This value represents the combined influence of the electric double layer (including Stern and diffuse layers) and the graphene quantum capacitance under the low-ionic-strength aqueous conditions of these experiments.This value lies within the expected range for electrolyte-gated graphene devices. It represents the combined influence of the electric double layer (including Stern and diffuse layers) and the graphene quantum capacitance under the low-ionic-strength aqueous conditions of these experiments. Because the commercial 20% HSA stock solution contains physiological saline that is progressively diluted with ultrapure water, the ionic strength varies with concentration. Under these conditions, the fixed $C_{eff}$ should be interpreted as an effective modeling parameter rather than a directly measured Debye screening length. The derivation and physical justification of the effective capacitance are provided in the Supplementary Information (Section S3).

Data were fitted in OriginPro 2025 by minimizing the sum of squared residuals between the experimental and model curves. Adjusted $R^2$ (Coefficient of Determination) $\geq$0.99 for all datasets, confirming the validity of the fitting approach across the concentration series. The extracted transport parameters were subsequently used to compute the inverse mobility signal ($S\mu$), which serves as the primary analytical metric for calibration, limit-of-detection analysis, and signal-to-noise evaluation presented in the Supporting Information (Section S4,S5, and S6).

### 2.5 Device Stability and Measurement Conditions

To minimize environmental drift and environmental noise, all measurements were performed in a low-noise, electrically grounded enclosure under ambient temperature and humidity. Each measurement sequence was initiated immediately after droplet equilibration to mitigate electrolyte evaporation and potential concentration changes. Devices were operated at sub-volt gate bias to prevent electrochemical reactions at the graphene–electrolyte interface. These precautions yielded excellent repeatability, minimal hysteresis, and stable baselines over multiple measurement cycles, validating the robustness of the graphene EGFET platform for precise albumin sensing.

## 3. Results and Discussion

### 3.1 Device Configuration and Electrostatic Operating Principle

The sensing architecture employs a graphene electrolyte-gated field-effect transistor (EGFET), as illustrated in Figure 1. The device integrates a monolayer graphene channel contacted by Au source and drain electrodes on a $Si/SiO_2$ substrate, while gating is achieved through an electrolyte droplet positioned directly above the active region. Unlike conventional solid dielectric gating, modulation of the channel occurs through ionic rearrangement[37] at the graphene-electrolyte interface, leading to formation of an electric double layer (EDL) with nanometer-scale charge separation When a gate potential is applied, counter-ions in the electrolyte accumulate near the graphene surface, forming an electric double layer that enables strong capacitive coupling between the gate and the channel. In the present work, serial dilutions of a commercial HSA stock solution in ultrapure water produce a low-ionic-strength aqueous environment in which the effective Debye screening length increases as ionic strength decreases. Under these conditions, surface-bound albumin molecules reside within the electrostatic interaction zone of the EDL and modulate the local carrier distribution and transport in the graphene channel without requiring direct charge transfer or chemical bonding. In this low-ionic-strength aqueous environment, the interfacial electric double layer exhibits enhanced capacitive coupling[35], enabling efficient modulation of carrier density at sub-volt gate bias.

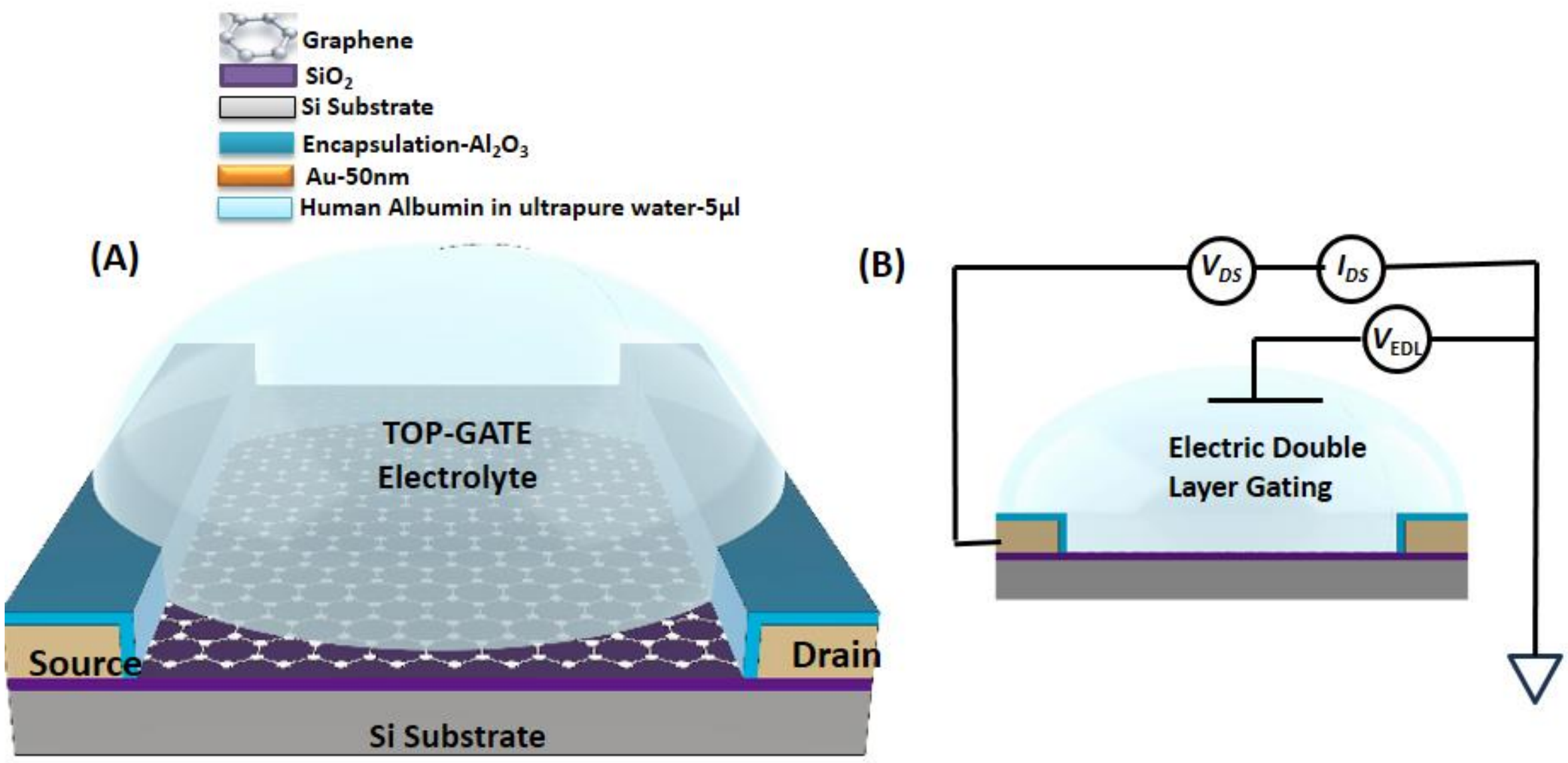

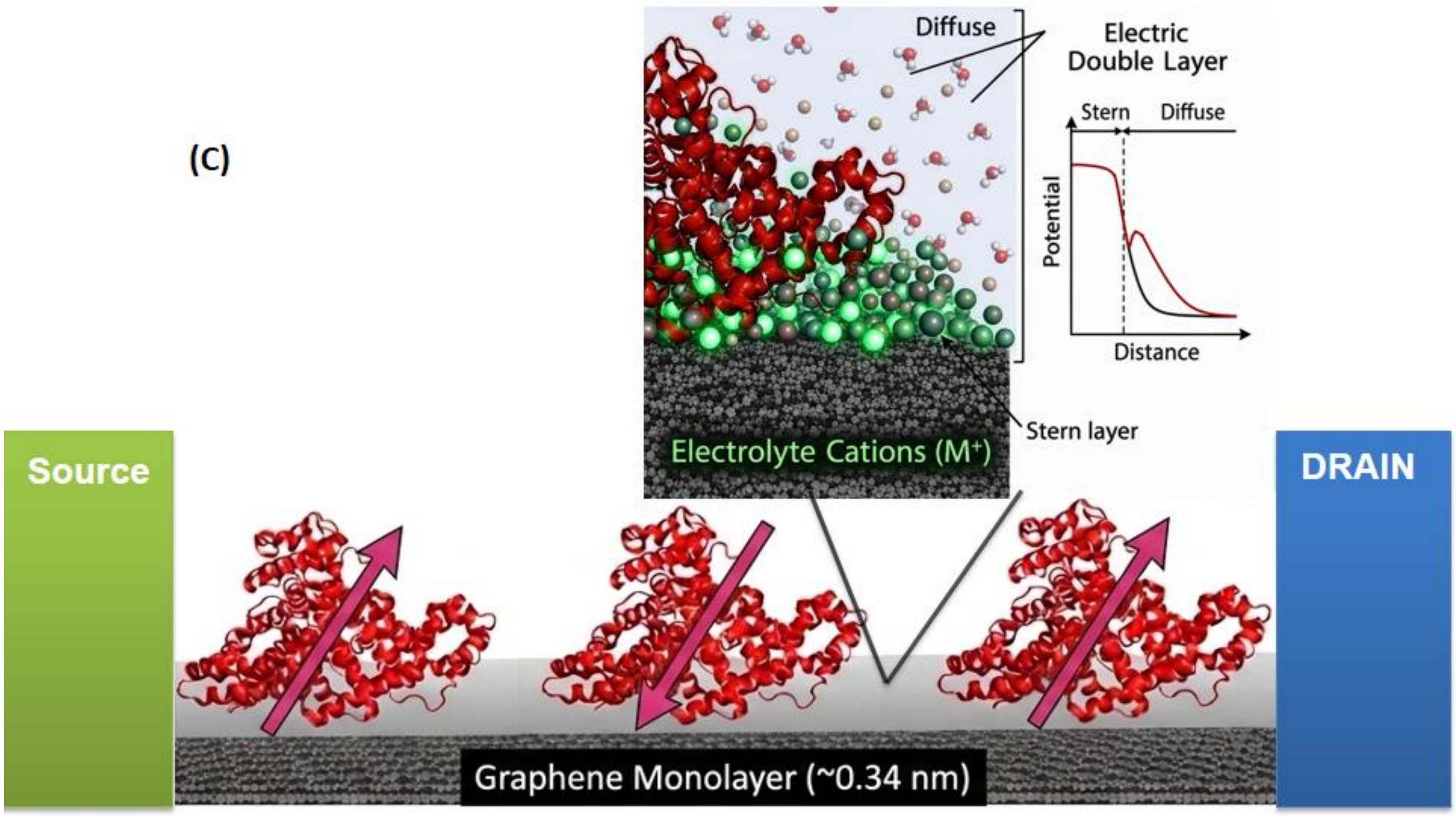

**Figure 1.** Liquid-gated graphene electric double-layer transistor (GEDLT) for human serum albumin (HSA) sensing. (A) Schematic representation of the monolayer graphene device architecture. (B) Electrolyte-gated measurement configuration illustrating electric double-layer (EDL) formation at the graphene–electrolyte interface. (C) Schematic illustration of the electrostatic sensing mechanism. Upon HSA adsorption, redistribution of electrolyte ions occurs within the EDL. Counter-ions ($M^+$) accumulate in the Stern layer adjacent to the graphene surface, while additional ions remain distributed within the diffuse layer. The resulting interfacial potential modulates the carrier density of the graphene monolayer and consequently alters the channel conductance under non-Faradaic sensing conditions.

For quantitative transport analysis, an effective gate capacitance was employed as a fitting parameter to capture the combined influence of the electric double layer (EDL) and the inherent quantum capacitance[36] of graphene. This effective parameter provides a consistent basis for extracting transport parameters such as carrier mobility and residual carrier density from transfer characteristics. While variations in ionic strength may lead to changes in the Debye length and interfacial capacitance, these effects are incorporated into the effective parameter. Treating '$C_{eff}$' as fixed therefore enables consistent comparison of extracted mobility trends across the concentration series. This approach enables consistent comparison of extracted mobility trends, which primarily reflect intrinsic transport modulation arising from adsorption-induced electrostatic perturbation.

Atomistic BD simulations reveal that HSA protein adsorb onto graphene surface through different binding orientations formed during diffusion toward the surface. The three most populated orientations (A–C), shown in Fig. 2, represent all sampled configurations. This orientational diversity is consistent with the non-polar nature of graphene, where interactions are dominated by Lennard-Jones terms rather than strong directional electrostatics, resulting in several energetically accessible adsorption poses rather than a single preferred geometry. The main interaction energy terms, cluster populations, and contact residues are summarized in Table S1 where contact residues are defined as those within 3.5 Å of the surface.

Among these configurations, Complex A is the most populated (50%) and exhibits the strongest interaction energy (−110.3 kT), dominated by van der Waals (i.e. Lennard-Jones) stabilization. Binding occurs mainly through residues in the central and C-terminal regions (e.g., ALA172, GLU297, LYS389, GLU396–GLN397, PRO441–ARG445), including several polar and charged amino acids at the interface (Figure 2A and Tab. S1). Complex B (43%) shows a slightly weaker interaction (−97.5 kT) but comparable van der Waals contributions, with contact residues distributed over broader sequence regions (ARG81–GLN94 and ALA362–VAL498), again enriched in polar and charged groups (Figure 2B and Tab. S1). Complex C (7%) represents the least favorable configuration (−93.9 kT), characterized by reduced Lennard-Jones contributions and more localized contact regions, which include electrostatically active residues (THR125–LYS174) (Figure 2C and Tab. S1).

Overall, these results indicate that adsorption is not governed by a unique binding motif, but mainly by a couple of configurations stabilized by van der Waals contact interactions, with electrostatics modulating the interfacial composition and orientation.

This molecular-level picture provides a direct framework for interpreting the device response. Adsorption of HSA onto graphene modifies the local electrostatic environment in two complementary ways: (i) by shifting the charge neutrality point through carrier-density modulation induced by the protein's interfacial charge distribution (Fig. S1), and (ii) by increasing spatial variability in the local potential due to the coexistence of multiple adsorption geometries and heterogeneous charge arrangements (Fig. 2).

Importantly, the combination of low orientational selectivity and strong interaction energies (reported in Tab. S1) helps explain the experimentally observed concentration dependence. As the HSA concentration increases, the surface becomes progressively covered by proteins adopting different orientations, leading to a thicker and more electrostatically heterogeneous interfacial layer. Although the simulations describe single-protein adsorption events, the favorable interaction energies indicate that adsorption is highly probable, supporting the formation of a dense protein layer at higher concentrations and thereby enhancing the overall electrostatic modulation of the graphene channel.

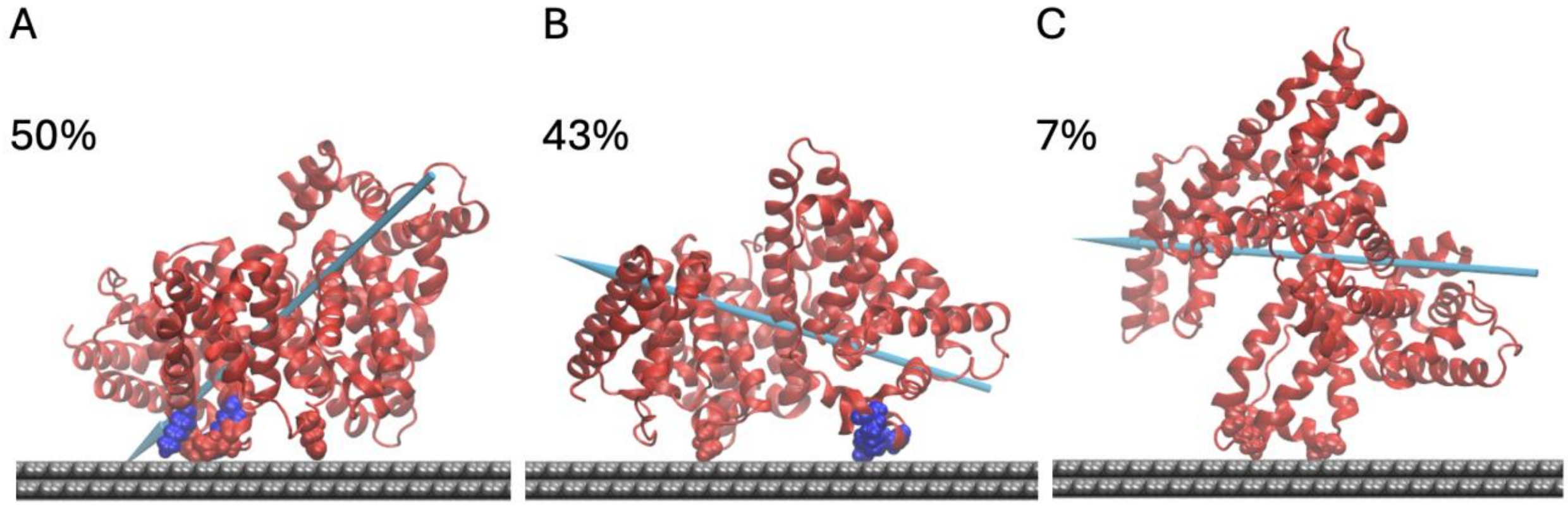

**Figure 2**. Representative orientations of the three most populated configurations of Human Serum Albumin adsorbed on a graphene surface following Brownian dynamics (BD) docking. Interfacial residues located within 3.5 Å of the surface are highlighted in van der Waals and colored according to their charge (positively charged: blue; negatively charged: red). The cyan arrow indicates protein dipole moment relative to the surface. The relative population of each orientation among all sampled configurations, is reported for the three most frequent adsorption geometries.

### 3.2 Output Characteristics and Channel-Dominated Transport Behavior

To examine the electrical behavior of the graphene channel under low-ionic-strength electrolyte gating, the output characteristics were first measured. Figure 3A shows the source–drain current '$I_{ds}$' as a function of drain voltage '$V_{ds}$' at zero gate bias for increasing HSA concentrations. Across the investigated range, the Ids/Vds curves remain linear and symmetric, indicating stable ohmic contact between graphene and the electrodes. No distortion or nonlinearity is observed, confirming that electrochemical processes or contact barriers do not dominate transport under the applied conditions.

Although linearity is preserved, the slope of the curves decreases progressively with increasing concentration. This corresponds to a reduction in output conductance, defined as $g_d = \partial Ids/\partial Vds$, extracted from the linear region of the characteristics. The evolution of $g_d$ with logarithmic HSA concentration is summarized in Figure 3B, showing a monotonic decline consistent with suppression of channel conductance. Since all measurements were performed under identical biasing conditions and in a non-Faradaic regime, the decrease in conductance is attributed to adsorption-induced modification of intrinsic graphene transport rather than to changes at the metal–graphene interface.

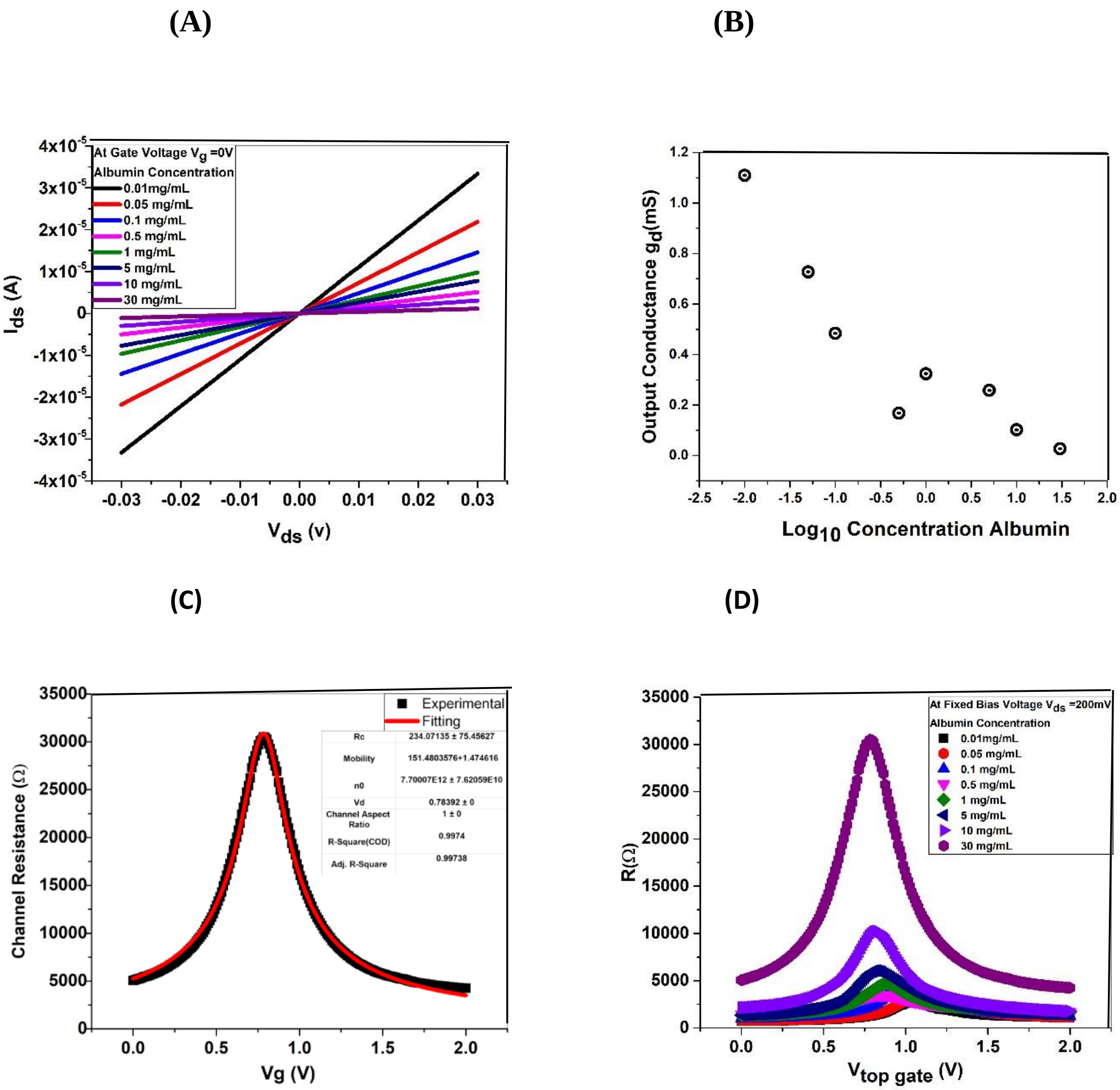

**Figure 3**. Output characteristics under increasing HSA concentration. (A) Source–drain current (Ids) as a function of drain voltage (Vds) measured at zero gate bias for increasing human serum albumin (HSA) concentrations, showing linear ohmic behavior.(B) Output conductance $g_d = \partial Ids/\partial Vds$ extracted from the linear region and plotted versus logarithmic HSA concentration.(C) Representative nonlinear fit of the resistance–gate voltage transfer characteristic used for extraction of transport parameters.(D) Transfer characteristics (channel resistance vs gate voltage) for increasing HSA concentrations, showing systematic Dirac voltage shift.

The gate-dependent transport was then analyzed through the transfer characteristics. A representative nonlinear fit of the resistance–gate voltage curve is presented in Figure 3C. The close agreement between experimental data and the model across both conduction branches confirms that the device behavior can be consistently described within the adopted transport framework. Following this validation, the same fitting procedure was applied to the transfer curves obtained at all investigated HSA concentrations, enabling systematic extraction of carrier mobility, residual carrier density, and contact resistance.

The transfer characteristics exhibit a well-defined ambipolar field-effect response, with a distinct resistance maximum corresponding to the charge neutrality (Dirac) point. A systematic displacement of the Dirac voltage is observed as the albumin concentration increases. The charge neutrality point shifts from 1.0452 V at 0.01 $mg.mL^{-1}$ to 0.7839 V at 30 $mg.mL^{-1}$, corresponding to an overall shift of approximately −261 mV. This shift corresponds to a change in carrier concentration of approximately $2.7\times10^{12}$ $cm^{-2}$, based on the effective gate capacitance of the electrolyte-gated configuration. Notably, while this level of electrostatic doping contributes to the observed Dirac shift, it is insufficient to account for the pronounced suppression of carrier mobility, indicating that disorder-induced scattering arising from adsorption-induced potential fluctuations plays a dominant role in the transport response, consistent with prior studies on disorder-limited transport in graphene[42]. Within the primary analytical range, the dependence of the Dirac voltage on logarithmic concentration is approximately linear. This behavior is consistent with electrostatic carrier-density modulation induced by negatively charged albumin molecules adsorbed within the Debye screening length.

The preserved linearity of the output characteristics together with the consistent fitting of the transfer curves confirms that the sensing response originates from intrinsic channel modulation within a stable non-Faradaic regime. While the Dirac voltage shift reflects overall carrier-density modulation induced by adsorbed albumin, the evolution of the transfer curves also suggests changes in the spatial uniformity of the electrostatic landscape. Such deviations from ideal field-effect behavior indicate increasing electrostatic inhomogeneity, which can influence carrier scattering beyond simple charge-density modulation. To clarify this effect, the following section examines the evolution of transconductance features and peak broadening as quantitative indicators of adsorption-induced disorder.

### 3.3 Quantification of Adsorption-Induced Disorder and Transport Evolution

Although the Dirac voltage displacement reflects global carrier-density modulation, the evolution of transfer characteristics reveals additional transport signatures that cannot be accounted for by electrostatic gating alone. The transconductance ($g_m$) characteristics were evaluated, as shown in Figure 4A, where the extrema corresponding to hole and electron conduction remain well defined across the investigated concentration range, confirming that the device retains stable field-effect functionality.

With increasing HSA concentration, however, the separation between the electron- and hole-branch extrema progressively expands. This peak-to-peak separation, denoted as ΔVpp (defined as the separation between transconductance extrema), is plotted in Figure 4B as a function of logarithmic concentration. The monotonic increase in ΔVpp indicates that the voltage window over which the channel transitions between dominant carrier types broadens with increasing surface coverage. This behavior indicates increasing electrostatic inhomogeneity within the graphene channel.

Further evidence of disorder development is obtained from analysis of the full width at half maximum (FWHM) of the resistance peak (Figure 4B), which provides a measure of the broadening of the charge neutrality region and the degree of carrier density inhomogeneity in the graphene channel. The FWHM increases systematically with concentration, indicating progressive broadening of the charge neutrality region. In graphene devices, broadening of the resistance maximum near the dirac point is commonly associated with spatial fluctuations in local carrier density, often described as charge puddle formation arising from disorder and electrostatic inhomogeneity in graphene[42,43,44].In the present case, these fluctuations arise from nonuniform electrostatic perturbations introduced by adsorbed albumin molecules distributed across the graphene surface. Such inhomogeneities enhance carrier scattering and contribute to the mobility reduction captured by the inverse mobility sensing metric Sμ. This behavior is consistent with an increase in the residual carrier density not used in the transport model, which captures disorder - induced charge inhomogeneity and provides a quantitative measure of the extent of charge puddle information.

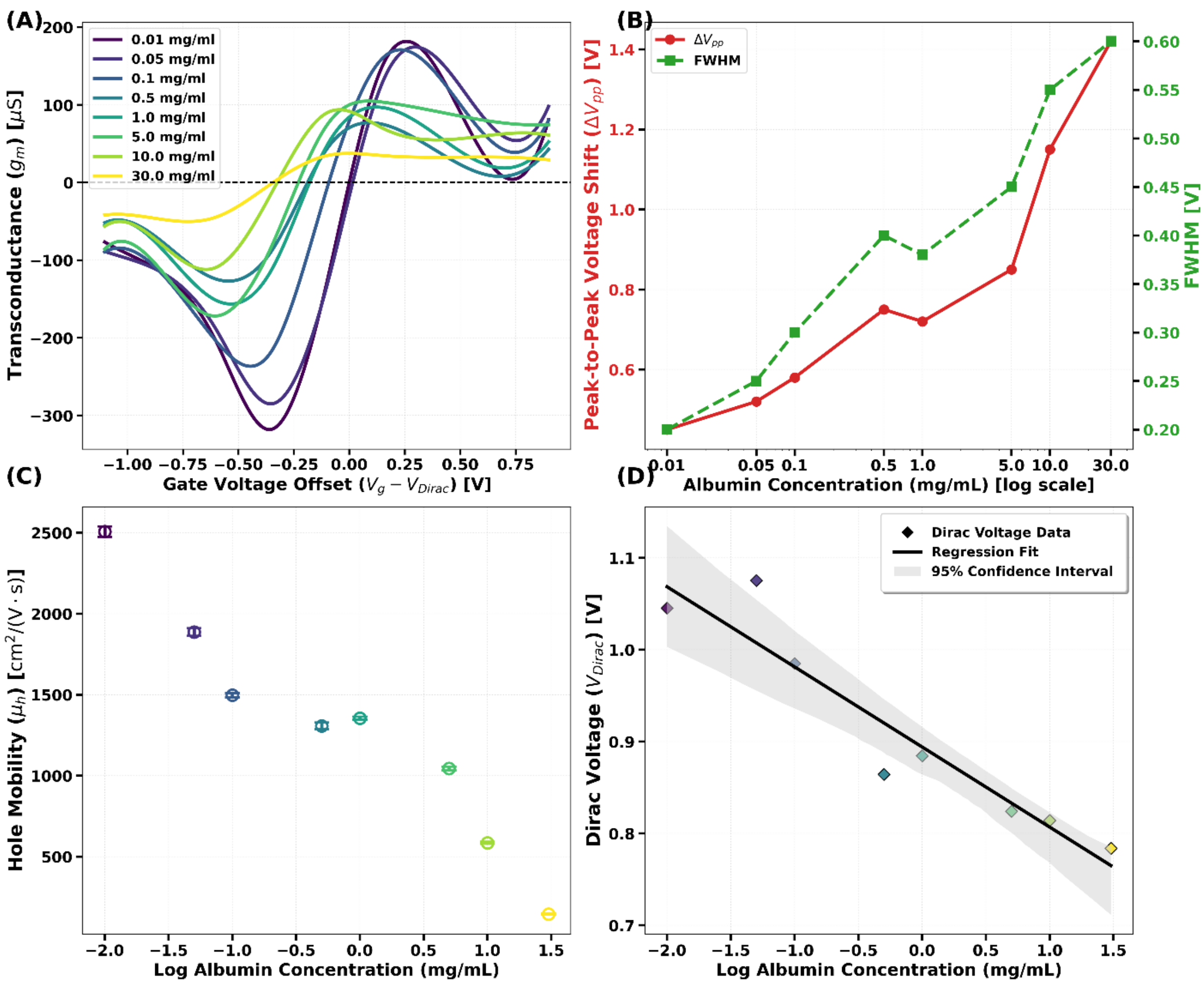

**Figure 4**. Electrical transport response of graphene devices as a function of HSA concentration.

(A) Transconductance (gm) as a function of gate voltage offset (Vg−VDirac)(B) Peak-to-peak voltage (ΔVpp) and full width at half maximum (FWHM) of the resistance peak versus concentration. ΔVpp is defined as the separation between transconductance extrema.(C) Extracted hole mobility (μh) showing systematic reduction with increasing concentration.(D) Dirac voltage shift as a function of concentration with linear regression and 95% confidence interval.

The concurrent increase in ΔVpp and FWHM indicates that adsorption induces not only a shift in the average carrier density but also a redistribution of the local potential landscape. While the Dirac shift captures the net electrostatic effect, ΔVpp and FWHM quantify the emergence of disorder-driven scattering. As surface coverage increases, these inhomogeneities intensify, thereby enhancing carrier scattering and ultimately reducing mobility[39].

To quantify the impact of adsorption-induced disorder on intrinsic transport, the transfer characteristics were fitted using a nonlinear resistance model that accounts for residual carrier density and field-effect modulation. From these fits, the corrected mobility (μ ) was extracted for each concentration and is summarized in Figure 4C.

A pronounced and systematic suppression of carrier mobility is observed with increasing HSA concentration. The hole mobility decreases from approximately 2506 $cm^2 V^{-1} s^{-1}$ at 0.01 mg $mL^{-1}$ to 151 $cm^2 V^{-1} s^{-1}$ at 30 mg $mL^{-1}$, corresponding to approximately an order-of-magnitude suppression of carrier transport. This reduction reflects increased carrier scattering resulting from adsorption-induced electrostatic fluctuations within the graphene channel. As surface coverage increases, the local potential landscape becomes increasingly nonuniform, shortening the effective carrier mean free path and reducing transport efficiency.

The evolution of the Dirac voltage ($V_{dirac}$) is shown in Figure 4D. A consistent negative shift is observed with increasing concentration, yielding an overall displacement of approximately −261 mV across the investigated range. The approximately log-linear dependence of $V_{Dirac}$ on concentration reflects electrostatic carrier-density modulation induced by surface-bound albumin molecules. The magnitude of mobility suppression exceeds that expected from carrier-density modulation alone, establishing disorder-enhanced scattering as the dominant mechanism governing the sensing response in the graphene channel. Building on this transport-based understanding of the sensing mechanism, the analytical performance of the device is evaluated in terms of sensitivity and limit of detection, as discussed in the following section.

### 3.4 Analytical Calibration, Signal-to-Noise Ratio, and Limit of Detection

The sensing response is quantified using the inverse mobility signal $S\mu$, which provides enhanced sensitivity to adsorption-induced disorder compared to direct mobility analysis.

Within the ultra-trace regime (0.01–0.1 mg.mL-1), the inverse mobility signal displays an approximately linear dependence on logarithmic concentration, yielding a coefficient of determination $R^2 \approx 0.957$. The lowest measured concentration (0.01 mg $mL^{-1}$) is used as a reference for baseline estimation, as no true zero-concentration measurement was performed. The linearity in this regime indicates consistent electrostatic coupling under low coverage conditions, where adsorption events remain sufficiently separated to avoid cooperative interactions.

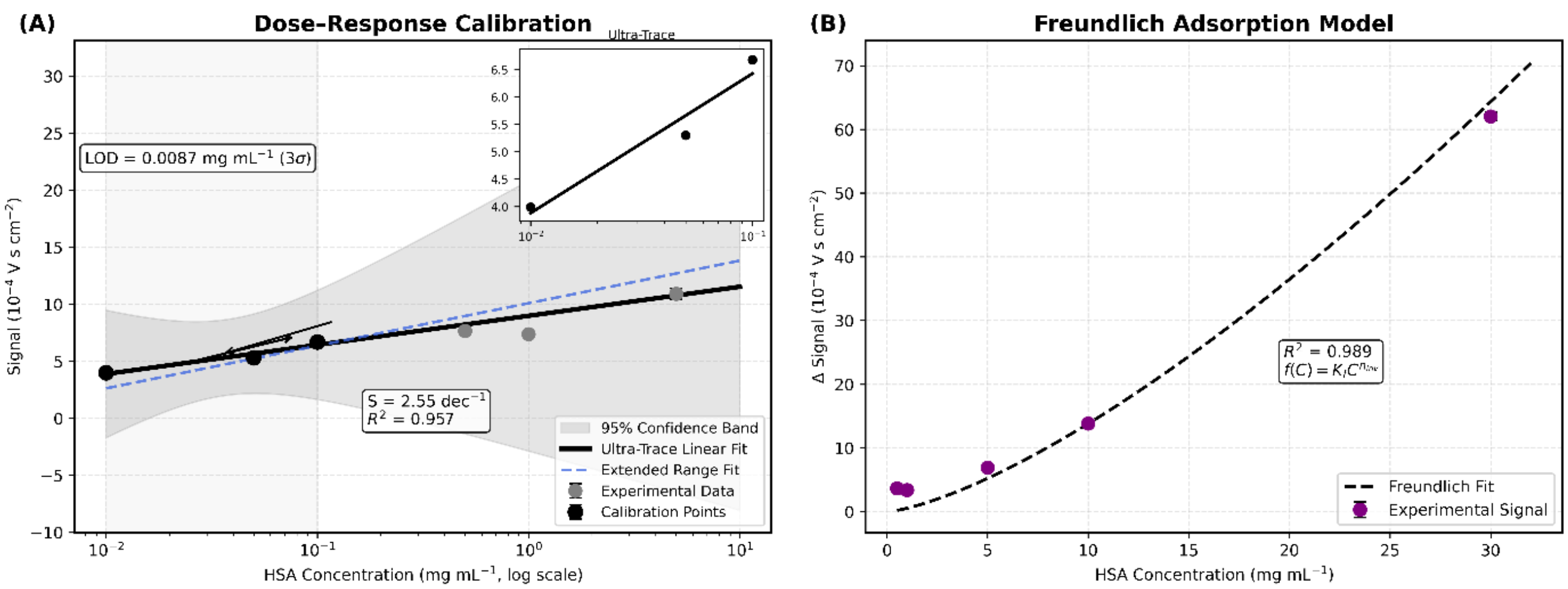

**Figure 5**. Dose–response analysis of human serum albumin detection using graphene EGFET. (A) Calibration curve of the inverse mobility signal Sμ as a function of logarithmic HSA concentration showing both the ultra-trace linear regime and the extended dynamic range. The inset highlights the ultra-trace region where the response exhibits high linearity. (B) Freundlich adsorption model describing the nonlinear response at higher concentrations.

The limit of detection (LOD) was estimated using the standard 3σ criterion, defined as LOD=3σ/S, where σ represents the standard deviation of the baseline signal and S is the slope of the calibration curve. Based on this approach, the LOD in the ultra-trace regime was determined to be 0.0087mg.mL$^{-1}$. It should be noted that this value lies below the lowest experimentally measured concentration and therefore reflects extrapolation of the linear calibration regime rather than direct measurement. A detailed derivation of the signal-to-noise ratio and limit-of-detection calculation is provided in the Supplementary Information (Section S6).

The low detection threshold reflects the enhanced electrostatic coupling enabled by reduced ionic strength, which increases the effective sensing distance within the Debye layer. Although ionic strength varies across the concentration series, several observations indicate that the measured response cannot be explained solely by electrolyte screening effects: (i) systematic mobility suppression exceeds that expected from electrostatic doping alone, (ii) disorder-related broadening parameters (ΔVpp and FWHM) evolve consistently with adsorption-induced charge inhomogeneity, and (iii) the BD simulations independently support formation of heterogeneous protein adsorption configurations capable of perturbing the local electrostatic landscape.The detection limit is consistent with previously reported graphene-based biosensors for protein detection, which typically achieve sensitivities in the microgram-per-milliliter range depending on device architecture and measurement conditions

Across the broader analytical window (0.01–10 mg $mL^{-1}$), the log-linear dependence is preserved with slightly reduced correlation ($R^2 \approx 0.76$), consistent with the onset of increased surface occupancy and spatial inhomogeneity. In this range, the calculated detection limit is 0.02mg.$mL^{-1}$. The modest decrease in linearity is attributable to the progressive development of disorder, as evidenced by the earlier increases in ΔVpp and FWHM.

At higher concentrations, deviation from strict log-linearity becomes more pronounced. Fitting this region using a Freundlich adsorption model provides excellent agreement ($R^2 \approx 0.989$), with an exponent $\frac{1}{n}$ >1 as shown in Figure 5B , consistent with cooperative or heterogeneous adsorption behavior. The use of the Freundlich model rather than the classical Langmuir isotherm is physically justified in the present system for several reasons. At elevated albumin concentrations, deviation from strict log-linearity indicates the onset of surface heterogeneity and cooperative adsorption effects. The classical Langmuir isotherm assumes a homogeneous surface with identical adsorption sites, monolayer coverage, and no lateral interaction between adsorbed species. However, protein adsorption on pristine graphene is governed by long-range electrostatic interactions within the electric double layer and is inherently heterogeneous due to spatial charge fluctuations, surface defects, and adsorbate-induced dipolar disorder. Furthermore, at high surface coverage, albumin–albumin interactions can modify the local electrostatic landscape, leading to cooperative effects that violate Langmuir assumptions. As discussed in comprehensive analyses of adsorption modeling, the Langmuir framework can be inappropriate for protein adsorption on heterogeneous substrates, particularly when conformational flexibility and intermolecular interactions are involved[40]. In contrast, the Freundlich isotherm accounts for energetically non-uniform surfaces and interaction-driven adsorption behavior, making it more suitable for describing the observed high-concentration regime[41]. The excellent agreement obtained with the Freundlich fit ($R^2 \approx 0.989$) therefore supports a heterogeneous, disorder-mediated adsorption mechanism rather than ideal monolayer saturation. The signal-to-noise ratio (SNR), defined as

$$SNR = \frac{\Delta S\mu}{\sigma}$$

where $\Delta S\mu = S\mu - S\mu_{,baseline}$ represents the change in inverse mobility signal relative to the baseline, and $S\mu$, baseline corresponds to the signal at the lowest measured concentration (0.01 mg.mL$^{-1}$) and σ is the standard deviation of the baseline signal used in the LOD calculation. The SNR increases systematically with concentration and exceeds unity even within the ultra-trace regime. This scaling confirms that mobility-based sensing enables reliable discrimination across both low- and clinically relevant-concentration domains. Importantly, the analytical response arises not solely from carrier-density modulation but also from the combined effects of electrostatic gating and disorder-enhanced scattering. By employing inverse mobility as the primary metric, the liquid-gated graphene EGFET achieves sensitive detection over multiple concentration regimes while maintaining non-Faradaic, reversible operation. The present study focuses on establishing the electrostatic and transport mechanism of adsorption-induced sensing in a controlled environment rather than demonstrating molecular specificity. Future studies will incorporate antifouling coatings and selective biofunctionalization strategies for operation in complex biological media.

Unlike conventional graphene biosensors, which primarily rely on Dirac voltage shifts as the sensing metric for biomolecule detection[17, 45, 46], the present approach exploits adsorption-induced disorder and carrier-scattering modulation quantified through the inverse mobility signal, enabling enhanced sensitivity and a broader analytical dynamic range. This transformation expands the dynamic range and emphasizes the contribution of disorder-induced scattering to the overall response. The magnitude of mobility suppression exceeds that expected from carrier-density modulation alone, establishing disorder-enhanced scattering as the dominant mechanism governing the sensing response in the graphene channel. The low detection limit arises from the strong sensitivity of carrier mobility to adsorption-induced disorder, where small perturbations in the local electrostatic landscape produce measurable changes in transport.

## Conclusion

In this study, a pristine graphene electrolyte-gated field-effect transistor (EGFET) was developed and systematically evaluated as a label-free platform for the quantitative detection of HSA. The device demonstrated a stable non-Faradaic operating regime in which albumin adsorption at the graphene–electrolyte interface modulated the channel conductance through electrostatic coupling mediated by the electric double layer (EDL). The concentration-dependent evolution of the output and transfer characteristics, including well-defined Dirac-point shifts and reversible changes in channel conductance, confirmed that the sensing mechanism is governed by long-range electrostatic coupling rather than chemical charge transfer or structural modification.

Complementary atomistic Brownian Dynamics (BD) simulations revealed that HSA adsorbs onto graphene through different binding orientations, characterized by heterogeneous interfacial charge distributions and variable dipole alignments relative to the surface, thereby providing a molecular-level basis for the observed electrostatic modulation and device response. Importantly, these simulations clarify that van der Waals interactions drive stable adsorption while electrostatics governs interfacial organization, offering a consistent physical explanation for both the Dirac-point shifts and the disorder-induced carrier scattering observed experimentally.

Comprehensive parameter extraction using the Fitting Transfer Model revealed that HSA adsorption induces systematic reductions in carrier mobility and broadening of the Dirac region. These trends are consistent with dipole-induced potential fluctuations and Coulomb scattering within the EDL, indicating that the protein layer perturbs the interfacial electronic landscape while maintaining stable field-effect behavior. The analytical calibration based on the inverse mobility metric further established the sensor's analytical performance, exhibiting a low detection threshold, a broad linear dynamic range spanning more than two orders of magnitude, and a well-defined saturation regime driven by Debye screening and surface-crowding effects.

Together, these results demonstrate that the reported graphene EGFET integrates high sensitivity, stable electrostatic transduction, and quantitative reliability, offering a promising platform for albumin detection across clinically relevant concentrations. The device's mechanistic clarity, structural simplicity, and reproducible performance highlight its potential for real-time protein monitoring in biomedical applications, including point-of-care diagnostics and dialysis-effluent assessment. Future efforts may extend this approach to multiplexed biomarker detection and clinical sample validation, further advancing graphene-based EGFETs toward practical diagnostic deployment.

Beyond albumin detection, the disorder-mediated sensing mechanism demonstrated here provides a new analytical framework for graphene-based biosensors. By exploiting adsorption-induced carrier scattering rather than relying solely on Dirac voltage shifts, the inverse-mobility approach enables enhanced signal amplification and improved robustness against baseline drift. This concept may be broadly applicable to the detection of other charged biomolecules and proteins in electrolyte-gated graphene platforms, opening new opportunities for scalable, low-power, and multiplexable bioelectronic sensing technologies.

The sensor demonstrated a wide dynamic range from 0.01 to 30 $mg.mL^{-1}$, successfully encompassing near- and sub-physiological concentrations of human serum albumin (HSA). This detection window is clinically significant for clinical real-time monitoring and for the early diagnosis of pathological hypoalbuminemia, commonly observed in chronic inflammation, hepatic cirrhosis, and nephrotic syndrome. Furthermore, the demonstrated limit of detection of 0.0087 mg/mL enables the quantification of trace albuminuria, offering perspective for a versatile platform for monitoring systemic protein loss across both serum and urinary matrices.

# Supporting Information

## Graphene Electric Double-Layer Transistors for Enhanced-Sensitivity Label-Free Detection of Human Serum Albumin

Arslan Liaquat[1], Ghasseem Baridi[2], Federico Rapuzzi[1], Daniele Goldoni[2], Vito Clericò[3], El Hadj Abidi[3], Yahya Moubarak Meziani[3], Mario Amado[3], Enrique Diez[3],Naveen Kumar[1], Camilia Coletti[4], Beatrice Cipriani[5], Hender Lopez[5], Giorgia Brancolini[6], Leonardo Martini[1], Luigi Rovati[2], Francesco Rossella[1*]

[1] Dipartimento di Scienze Fisiche, Informatiche e Matematiche, University of Modena and Reggio Emilia, Via Campi 213/A, 41125 Modena, Italy

[2] Department of Engineering Enzo Ferrari, University of Modena and Reggio Emilia, via Vignolese 905, 41125 Modena, Italy

[3]USAL–Nanolab, Departamento de Física Fundamental University of Salamanca Salamanca 37008, Spain

[4] Center for Nanotechnology Innovation @NEST, Istituto Italiano di Tecnologia, Piazza San Silvestro 12, I-56127 Pisa, Italy

[5] School of Physics, Clinical and Optometric Sciences, Technological University Dublin, Grangegorman, Ireland

[6] Institute Nanoscience, CNR, S3, Via G. Campi 213/A, Modena

Corresponding author: Arslan Liaquat, arslan.liaquat@unimore.it; Francesco Rossella, francesco.rossella@unimore.it

**S1. Brownian Dynamics Docking Analysis of HSA on Graphene**

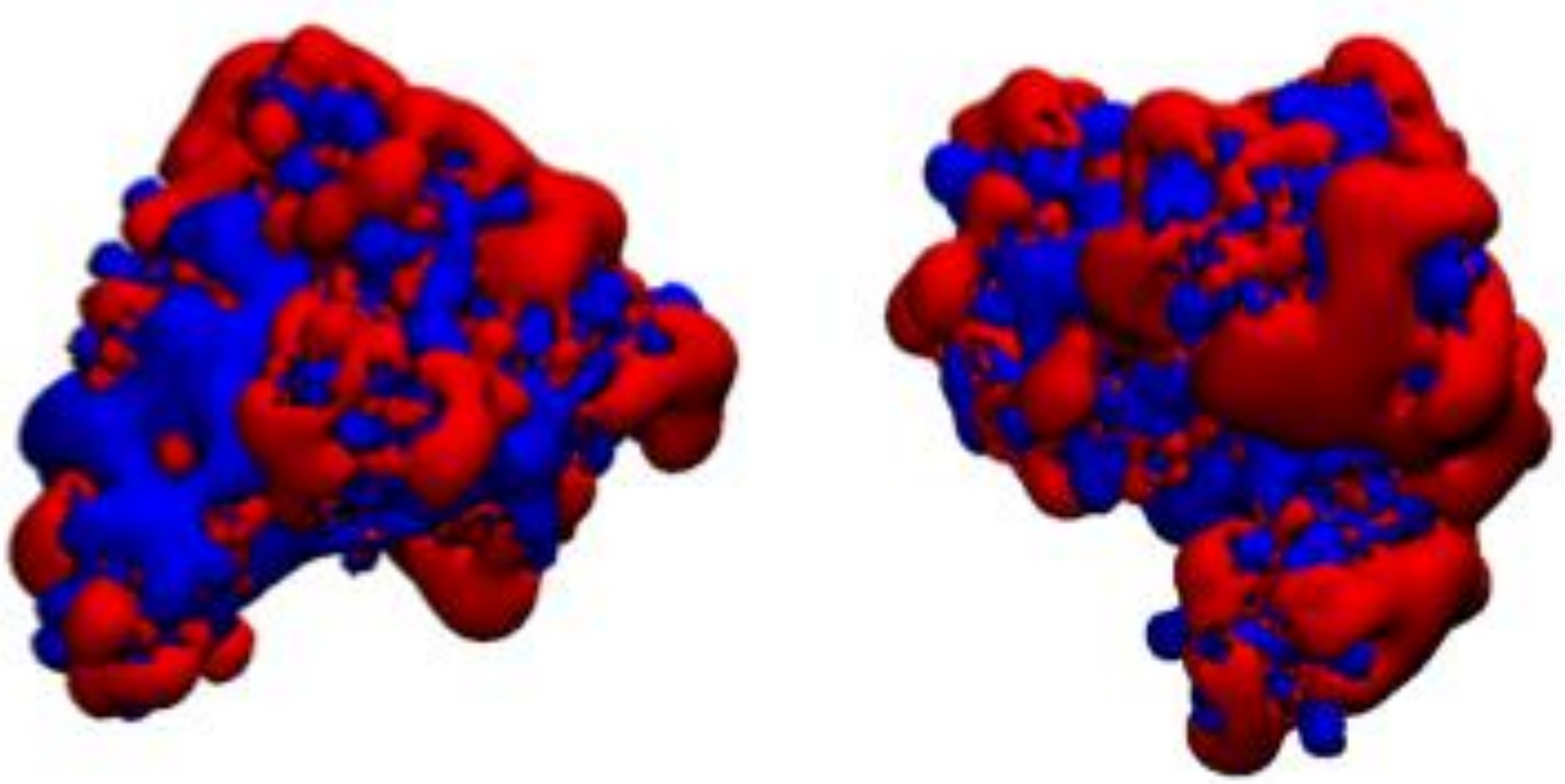

**Fig. S1** Electrostatic potential map of Human Serum Albumin illustrating the spatial distribution of charges on the protein surface. Negative regions are shown in red, while positive regions are shown in blue

Brownian Dynamics docking simulations were performed to investigate the preferred adsorption configurations of HSA on graphene and to evaluate the resulting electrostatic heterogeneity at the interface. The simulations reveal multiple energetically favorable adsorption orientations associated with distinct charge distributions and surface-contact residues. Representative docking clusters and their interaction-energy components are summarized in Table S1.

| **Label** | **$U_{tot}$[a]** kT | **$U_{des}$[b]** kT | **$U_h$[c]** kT | **LJ[d]** kT | **Cluster size[e]** | **Contact residues (<3.5 A)[f]** |
|---|---|---|---|---|---|---|
| **A** | -110.3 | 46.2 | -14.6 | -141.9 | 50% | ALA172 **GLU297** **LYS389** **GLU396** GLN397 PRO441 **GLU442** **ARG445** |
| **B** | -97.5 | 58.8 | -14.7 | -141.6 | 43% | **ARG81** ALA92 **LYS93** GLN94 ALA362 ALA364 **ASP471** THR496 VAL498 |
| **C** | -93.9 | 14.9 | -7.8 | -101.1 | 7% | THR125 HID128 **ASP129** GLN170 ALA172 **LYS174** |

**Table S1** Complexes resulting from BD docking between HSA and graphene surface

a: total interaction energy of cluster in kT with T = 300 K

b: surface desolvation energy, in kT

c: nonpolar (hydrophobic) desolvation energy, in kT

d : Lennard-Jones energy term

e: Relative population of the selected clusters reported in percentage.

f: Residues with atoms contacting graphene at distances < 3.5 Å. In red negatively charged residues and in light blue positively charged residues.

**S2. Device Configuration and Electrostatic Gating:**

All electrical measurements were performed using commercially fabricated Graphenea mGFET-4D chips operated in a droplet-based liquid-gating configuration with ultrapure water (18.2 MΩ·cm) as the electrolyte. In this geometry, gating occurs through formation of the electric double layer (EDL) at the graphene/electrolyte interface rather than through the underlying $SiO_2$ dielectric.

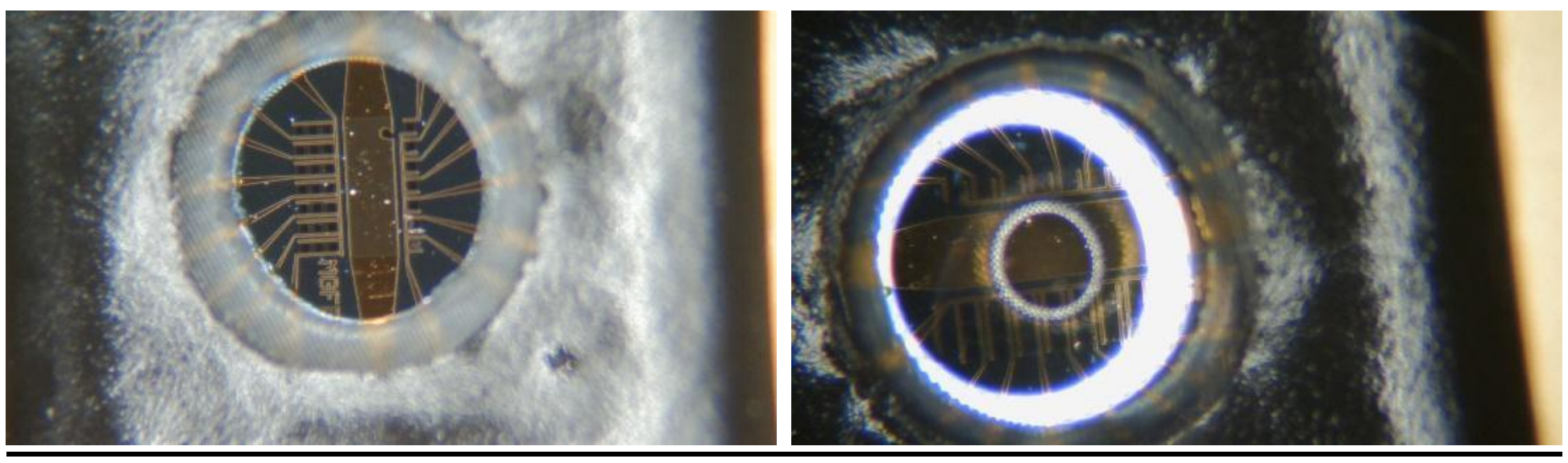

**Fig. S2** Schematic illustration of the graphene electrolyte-gated field-effect transistor (EGFET) configuration used for HSA sensing measurements. Electrical gating is achieved through formation of the electric double layer at the graphene/electrolyte interface under droplet-based liquid-gating conditions.

Under these low-ionic-strength conditions, the Debye screening length remains relatively large and varies moderately due to residual ionic species introduced during dilution of the commercial HSA stock solution. Consequently, the interfacial capacitance is not strictly constant across the concentration series. To ensure consistent extraction of transport parameters, an effective gate capacitance was introduced as a fixed fitting parameter within the transport model

**S3. Effective Capacitance Model and Justification**

The effective gate capacitance $C_{eff}$ accounts for the series combination of the electric double-layer capacitance $C_{EDL}$ and the graphene quantum capacitance $C_q$:

$$\frac{1}{C_{eff}} = \frac{1}{C_{EDL}} + \frac{1}{C_q}$$

$$C = \frac{\epsilon\epsilon_O}{\lambda_D}$$

Where $\lambda_D$ is the Debye screening length, $\varepsilon$ is the relative permittivity, and $\varepsilon o$ is the vacuum permittivity.

Under low-ionic-strength aqueous conditions, $\lambda_D$ is on the order of several nanometers, resulting in $C_{EDL}$ in the range of ~1–3 μFcm$^{-2}$. Near the Dirac point, the graphene quantum capacitance becomes comparable in magnitude, leading to an effective series capacitance in the range of approximately 1–2 μF cm$^{-2}$.

Based on this physically expected range, a constant value of $C_{eff}$=1.65 μF cm$^{-2}$ was adopted for all fitting procedures.

This value was selected to ensure stable and physically consistent extraction of transport parameters across the full concentration range. The parameter should therefore be interpreted as an effective modeling capacitance rather than a directly measured electrostatic quantity.

Although ionic strength varies moderately across the concentration series due to dilution of the saline-containing HSA stock solution, the observed transport evolution cannot be explained solely by electrolyte screening effects. In particular, the systematic suppression of carrier mobility and the evolution of disorder-related broadening parameters support adsorption-induced electrostatic inhomogeneity within the graphene channel.

**S4. <u>Transport Model and Parameter Extraction</u>**

The total device resistance was modeled using the standard graphene transport expression:

$$R = Rc + \frac{L}{W} \cdot \frac{1}{\mu\, C_{eff} \sqrt{no^2 + (Vg - Vdirac)^2}}$$

Where Rc is the contact resistance, L and W are the channel length and width , respectively. μ is the carrier mobility, Ceff is the effective gate capacitance, n0 is the residual carrier densityVg is the applied gate voltage and $V_{Dirac}$ is the charge neutrality voltage

For all the devices analyzed the channel aspect ratio (L/W=1). The key transport parameters extracted from the nonlinear fitting of the transfer curves are summarized in the table S.

The Dirac voltage provides information on charge neutrality shifts induced by protein adsorption, while the extracted hole mobility reflects changes in carrier scattering within the graphene channel. The inverse mobility is used as the sensing signal because adsorption-induced disorder primarily modifies carrier mobility rather than carrier density

The extracted Dirac voltage reflects charge-neutrality shifts induced by adsorption of negatively charged HSA molecules, whereas the carrier mobility provides information on adsorption-induced carrier scattering and electrostatic disorder within the graphene channel. The inverse mobility signal, $S\mu=1/\mu h$ was used as the primary sensing metric because adsorption-induced disorder predominantly affects carrier mobility rather than carrier density alone.

**Table S2.** Extracted transport parameters obtained from nonlinear fitting of graphene EGFET transfer characteristics at different HSA concentrations.

| HSA conc. (mg/mL) | Log ( Conc.) | Dirac Voltage (V) | Mobilitty | Inverse Mobility ( $10^{-4}$ ) | $R^2$ |
|---|---|---|---|---|---|
| 0.01 | −2.0000 | 1.0452 | 2505.9 | 3.99 | 0.997 |
| 0.05 | −1.3010 | 1.0754 | 1885.9 | 5.3 | 0.997 |
| 0.1 | −1.0000 | 0.9849 | 1496.5 | 6.68 | 0.999 |
| 0.5 | −0.3010 | 0.8643 | 1306.5 | 7.65 | 0.998 |
| 1 | 0 | 0.8844 | 1355 | 7.38 | 0.999 |
| 5 | 0.699 | 0.8241 | 917.4 | 10.9 | 0.952 |
| 10 | 1 | 0.8141 | 560.6 | 17.84 | 0.995 |
| 30 | 1.4771 | 0.7839 | 151.5 | 66.01 | 0.99 |

## S5. Additional Transport Metrics and Device-Level Analysis

To further evaluate the transport response induced by HSA adsorption, additional device-level metrics were extracted from the fitted transfer characteristics, including charge-neutrality broadening, contact resistance evolution, and intrinsic transconductance behavior.

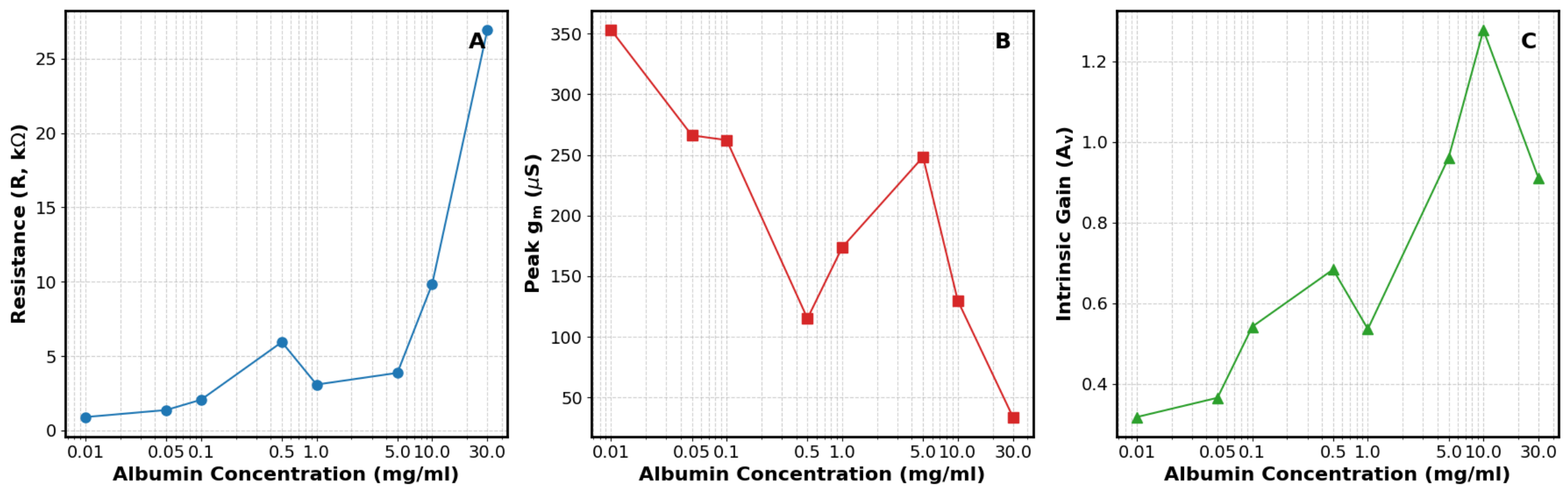

**FigureS3**:**(A)** Channel resistance extracted from the linear transport regime, showing a progressive increase with albumin concentration due to adsorption-induced carrier scattering and reduced conductivity.**(B)** Peak transconductance ($g_m$) as a function of HSA concentration, exhibiting an overall reduction that reflects degradation of gate coupling efficiency and carrier mobility under increasing surface disorder.**(C)** Intrinsic gain ($A_v$) calculated from the ratio of transconductance to output conductance, illustrating the combined impact of protein adsorption on transconductance suppression and channel transport properties.

**Table S3: Extracted values rof Channel Resistance, Transconductance and intrinsic gain derived from Graphene EGFET transfer characteristi cs**

| Concentration (mg/mL) | Resistance R (Ω) | gm (µS) | Intrinsic Gain Av |
|---|---|---|---|
| 0.01 | 901 | 353 | 0.318 |
| 0.05 | 1,375 | 266.1 | 0.366 |
| 0.1 | 2,068 | 262.2 | 0.542 |
| 0.5 | 5,942 | 115.1 | 0.684 |
| 1 | 3,086 | 173.8 | 0.536 |
| 5 | 3,872 | 248.3 | 0.96 |
| 10 | 9,846 | 129.8 | 1.28 |
| 30 | 26,927 | 33.8 | 0.911 |

To evaluate the transistor-level response of the device during sensing, the channel resistance, peak transconductance, and intrinsic gain were extracted from the transfer characteristics at each albumin concentrationThe intrinsic gain represents the maximum small-signal amplification capability of the device and is defined as:

$$A_v = g_m \cdot R$$

where gm is the peak transconductance and R is the channel resistance at the chosen operating point.The resistance and transconductance values used for the calculation were extracted from the transfer characteristics at a fixed drain bias of 200 mV,. The transconductance values were converted from μS to S prior to multiplication.The intrinsic gain exhibits non-monotonic behavior across the investigated concentration range.At low concentrations (0.01–0.1 mg $mL^{-1}$), the gain remains modest despite high transconductance values because the channel resistance is relatively low. As concentration increases, adsorption-induced electrostatic disorder enhances channel resistance while maintaining sufficient transconductance, resulting in amplification of Av.The maximum intrinsic gain is observed at 10 mg $mL^{-1}$ (Av=1.280). At this intermediate regime, the balance between transconductance and resistance is optimized, yielding the highest small-signal amplification capability.At 30 mg $mL^{-1}$, intrinsic gain decreases despite large channel resistance. This reduction is attributed to significant mobility suppression and transconductance degradation caused by increased carrier scattering and partial obstruction of the graphene conduction pathway by the adsorbed protein layer.

The intrinsic gain behavior supports the disorder-mediated sensing mechanism. At moderate surface coverage, electrostatic inhomogeneity enhances channel resistance without fully suppressing transconductance, thereby increasing gain. At high coverage, however, excessive scattering reduces transconductance sufficiently to limit amplification.

This non-monotonic trend confirms that device performance is governed by the interplay between electrostatic gating and disorder-enhanced scattering rather than simple carrier-density modulation alone.

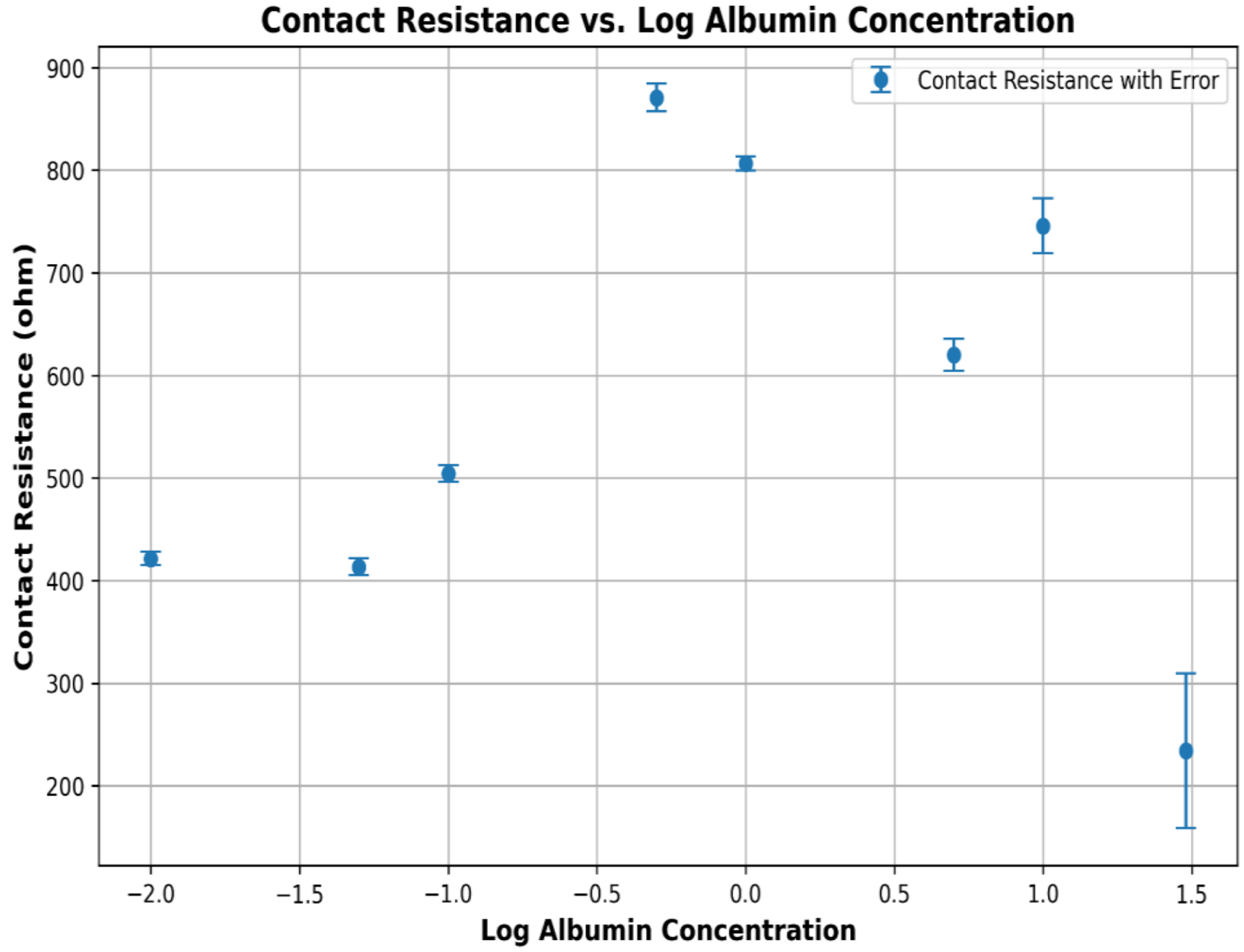

**Figure S4: Contact Resistance as function of Log.Albumin Concentration**

The extracted contact resistance exhibits non-monotonic behavior across the concentration range, reflecting the interplay between carrier density modulation and mobility reduction. These variations remain within fitting uncertainty and do not dominate the sensing response, confirming that the primary signal originates from intrinsic channel transport.

The responsivity increases with concentration, reflecting enhanced sensitivity of the inverse mobility signal to adsorption-induced disorder. The observed nonlinear increase is consistent with the transition from isolated adsorption events to cooperative and heterogeneous surface interactions at higher concentrations.

## S6. Ultra-Trace Calibration and Limit of Detection

In the ultra-trace regime, the sensing signal was defined as the inverse hole mobility:

$$S\mu = 1/\mu h$$

For plotting convenience, the signal was scaled as:

$S\mu = (1/\mu h) \times 10^4$ (units: $10^{-4}$ V·s·cm$^{-2}$)

The ultra-trace analytical range was defined as 0.01-0.1 mg mL$^{-1}$.

The signal-to-noise ratio (SNR) was calculated as:

$$SNR = \frac{\Delta S\mu}{\sigma baseline}$$

$S\mu = S\mu_i - S\mu_{0.01}$ and σbaseline is the propagated uncertainty of the baseline signal.

Slope = m = (6.68 − 3.99) / [log10(0.1) − log10(0.01)] = 2.69

In SI calibration units:
Slope = m = $2.69 \times 10^{-4}$ $V \cdot s \cdot cm^{-2}$ per decade

## Limit of Detection (LOD)

The propagated baseline uncertainty from the mobility fitting was:

$\sigma_{baseline} = \sigma_\mu / \mu_h{}^2 \times 10^4$

$\sigma_\mu$ = Mobililty Fitting Uncertinity

$\mu_h$ = Hole Mobility

$\sigma_{baseline} = 0.0518$ $V \cdot s \cdot cm^{-2}$ (For Conc. 0.01)

Using the 3σ/Slope criterion:

$LOD_{log}$ = (3σbaseline) / slope (m) = 0.058 decades

Since $Conc._{min}$ = 0.01 mg $mL^{-1}$ (log10 = –2):

log10 (LOD) = −2 − 0.058 = −2.058

LOD = $10^{-2.058} = 8.7 \times 10^{-3}$ mg $mL^{-1}$

LOD (ultra-trace regime) = 0.0087 mg $mL^{-1}$

.

**Table S4. Ultra-trace calibration data for HSA detection using the inverse-mobility sensing signal.**

| Conc. (mg/mL) | log10(Conc.) | μh (cm²/Vs) | Sμ(×10⁻⁴ Vs/cm²) | ΔS | SNR |
|---|---|---|---|---|---|
| 0.01 | -2 | 2505.93 | 3.99 | 0 | 0 |
| 0.05 | -1.301 | 1885.94 | 5.3 | 1.31 | 4.43 |
| 0.1 | -1 | 1496.47 | 6.68 | 2.69 | 9.09 |

**Table S5. Ultra-trace calibration data for HSA detection using the inverse-mobility sensing signal.**

| HSA conc. (mg/mL) | log10(C) | $\sigma_{\mu}$ (cm²/Vs) | $\mu_h^2$ | $\sigma_{baseline}$ | Signal | SNR | Rc (Ω) |
|---|---|---|---|---|---|---|---|
| 0.01 | -2 | 32.5 | $6.2797\times10^{6}$ | 0.0518 | 0 | 0 | 421.57 |
| 0.05 | -1.301 | 24.1 | $3.5568\times10^{6}$ | 0.0678 | 1.31 | 25.31 | 413.39 |
| 0.1 | -1 | 13.4 | $2.2394\times10^{6}$ | 0.0598 | 2.69 | 51.98 | 504.14 |
| 0.5 | -0.301 | 21.1 | $1.7068\times10^{6}$ | 0.1236 | 3.66 | 70.72 | 870.66 |
| 1 | 0 | 9.5 | $1.8361\times10^{6}$ | 0.0517 | 3.39 | 65.5 | 806.66 |
| 5 | 0.699 | 39.3 | ($8.4153\times10^{5}$ | 0.467 | 6.91 | 133.52 | 806.02 |
| 10 | 1 | 7.6 | ($3.1426\times10^{5}$ | 0.2418 | 13.85 | 267.61 | 830.93 |
| 30 | 1.4771 | 1.5 | $2.2943\times10^{4}$ | 0.6538 | 62.03 | 1198.55 | 234.07 |